# Sorting and separation of microparticles by surface properties using liquid crystal-enabled electro-osmosis

Chenhui Peng[a,1], Taras Turiv[a], Yubing Guo[a,2],

Qi-Huo Wei[a,b] and Oleg D. Lavrentovich[a,b,*]

[a]*Liquid Crystal Institute and Chemical Physics Interdisciplinary Program, Kent State Universtity, Kent, OH 44242*

[b]*Department of Physics, Kent State Universtity, Kent, OH 44242*

[1]*Current affiliation: Research Laboratory of Electronics, Massachusetts Institute of Technology, Cambridge, MA, 02139*

[2]*Current affiliation: Max Planck Institute for Intelligent Systems, Stuttgart, Germany 70569*

*Corresponding author: e-mail: olavrent@kent.edu, tel.: +1-330-672-4844.

**Abstract**

Sorting and separation of microparticles is a challenging problem of interdisciplinary nature. Existing technologies can differentiate microparticles by their bulk properties, such as size, density, electric polarizability, etc. The next level of challenge is to separate particles that show identical bulk properties and differ only in subtle surface features, such as functionalization with ligands. In this work, we propose a technique to sort and separate particles and fluid droplets that differ in surface properties. As a dispersive medium, we use a nematic liquid crystal (LC) rather than an isotropic fluid, which allows us to amplify the difference in surface properties through distinct perturbations of LC order around the dispersed particles. The particles are placed in a LC cell with spatially distorted molecular orientation subject to an alternating current electric field. The gradients of the molecular orientation perform two functions. First, elastic interactions between these pre-imposed gradients and distortions around the particles separate the particles with different surface properties in space. Second, these pre-imposed patterns create electro-osmotic flows powered by the electric field that transport the sorted particles to different locations thus separating them. The demonstrated unique sorting and separation capability opens opportunities in lab-on-a-chip, cell sorting and bio-sensing applications.

## 1. Introduction

Sorting and separation of microparticles is of prime importance in chemical and biological analysis, diagnostics, colloidal and molecular research [1-3]. The difficulty of sorting is caused by the inherently laminar nature of the flow in a microfluidic channel. Current particle-sorting techniques can be classified either as passive [2], based on embedded obstructions to laminar flows [4-7] or active, involving externally applied forces [1, 2], such as magnetic [8-10] and electric [3, 11-13] fields, optical [14-17] and acoustic waves [18, 19]. All these methods differentiate particles according to their bulk properties, such as size or mass (in microfluidic devices with obstacles [4-7]), refractive indices [14-17], density and compressibility (in acoustic sorting [18, 19]), magnetic susceptibilities [20], dielectric permittivity [13], electric conductivity [1-3], etc. The listed techniques bring innovation to various diagnostic and therapeutic applications but are not effective when the particles are similar in bulk properties and differ only in surface features. Sorting particles that differ only in surface properties is of importance in immunological applications, cell separation, nucleic acid research, in which colloidal particles such as silica, polymers, or metallic microspheres [21], are functionalized with ligands designed for a specific binding in order to trigger an antibody-antigen reaction of interest [22-25].

In this work, we demonstrate a method to sort and spatially separate particles that differ in surface properties, by using a pre-patterned nematic liquid crystal (LC) instead of a homogeneous isotropic fluid, as a dispersive medium. The main feature of a LC is its long-range orientational order described by the so-called director $\hat{\mathbf{n}} \equiv -\hat{\mathbf{n}}$. The director field is distorted around dispersed particles into different structures that depend on the surface properties of the particles (the so-called anchoring effect) [26]. These local distortions interact elastically with the background director field in the LC cell, moving the particles to the regions that minimize the combined surface anchoring and elastic energy [27-31]. The background distortions of the LC can even drive molecular self-assembly at nanoscale [32, 33].

If the colloids are forced to move, say, under the action of gravity [34] or electric field [35], these local director distortions can be used to steer the particles along certain trajectories. The control of trajectories can be achieved by designing a set of physical obstacles in the LC cell with their own surface anchoring properties, as demonstrated by Chen et al. [34], or by using patterned surface alignment of the LC [35, 36] without relying on physical obstacles.

In this work, to produce the effects of sorting and separation of differently anchored particles, we use liquid crystal-enabled electro-osmosis (LCEO) [37] in LC cells with patterned director fields [36]. The pre-patterned director field, created by photoalignment [38], plays a dual role. First, it creates a spatially-varying elastic potential energy profile that drives the colloidal particles into locations in which their surface properties match the type of director distortions [39]. Second, in presence of an applied alternating-current (AC) electric field, these director distortions activate electro-osmotic flows that move the particles of different surface properties towards the opposite ends of the sorting device. We first describe the idea of sorting and separation and then present experimental realizations for solid particles and droplets of fluids.

## 2. Materials and Methods

### *2.1. Photopatterning the photoalignment azo dye layer.*

The photosensitive material azo dye Brilliant Yellow (BY) is purchased from *Sigma* and used without further purification. BY is mixed with N,N-Dimethylformamide (DMF) solvent at 0.4 wt% concentration. Glass substrates are washed in ultrasonic bath with Cavi-clean detergent, then with isopropyl alcohol and dried in the oven at 80°C for 15 minutes. Subsequently the substrates are placed in ozone chamber under UV exposure for about 5 minutes. The BY solution is spin coated on the glass substrates at 3000 rpm for 30 seconds. The glass plates are then baked at 95ºC for 30 minutes. Two separate substrates are exposed for 5 minutes in the photo-patterning exposure system with the plasmonic metamask [38, 39]. The locally polarized beam causes realignment of the photosensitive BY molecules perpendicularly to the polarization of light. The pattern of light polarization is thus imprinted into the BY layer.

## *2.2. Preparation of colloidal suspension in LC*

The streptavidin-coated spheres are purchased from *Bangs Laboratory, INC*. DMOAP-coated and streptavidin-coated spheres are mixed with LC of 0.01 wt% respectively. The colloidal suspension in LC is then inject into the photopatterned cells by capillary force. The boomerang-shaped colloids are fabricated from the epoxy-based photoresist SU-8 following the procedure in Ref. [40], with symmetric arms of length $a = 2.1\,\mu m$ each, thickness $0.51\,\mu m$, width $0.55\,\mu m$, and an apex angle $100-110°$. The boomerang-shaped colloids and DMOAP-coated spheres are suspended with LC in the way discussed above and inject in the cells.

## *2.3. Preparation of droplet suspension in LC*

1 wt% of sodium dodecyl sulfate (SDS) purchased from *Sigma* is mixed with glycerol (*Sigma*) and then put in the oven with temperature 65ºC for 15 minutes. Afterwards, this mixture is vortexed for 2 minutes to make it homogeneously mixed. Glycerol with SDS and pure glycerol are mixed with LC at 0.01 wt% respectively and vortexed for 2 minutes to prepare both droplets existing in the suspension. The droplets suspension is injected into the photo-patterned cells by capillary action. 1,2-Dilauroyl-sn-glycero-3-phosphorylcholine (DLPC) is purchased from *Avanti Polar Lipids, Inc.* and mixed with glycerol in the way described above. The glycerol droplets with and without DLPC are suspended in the LC and injected into the cells.

## *2.4. Preparation of silicone oil droplet suspension in LC*

1 wt% silicone oil (*Sigma*) is mixed with polystyrene spheres of 0.01 wt% in the LC. This mixture is sonicated in the water bath of 75ºC for 2 minutes and then injected into the photo-patterned cells right after the sonication.

# 3. Results

## *3.1. Surface anchoring*

LCs are known for their extreme sensitivity to the chemical and physical features of adjacent surfaces [41]. Anisotropic molecular interactions between the LC and an adjacent medium set up a preferred orientation of the LC director that can be perpendicular (homeotropic) to the interface, tangential, or tilted. For example, addition of a small quantity of a surfactant changes the LC anchoring from tangential to homeotropic at interfaces with both solid [41] and fluid media [42].

Different types of surface anchoring reflect the chemical structure of the surface and sometimes more subtle features such as surface roughness, degree of coverage with self-assembled monolayers, etc. [43-45]. For example, numerical simulations by Claudio Zannoni's group demonstrated that the rod-like molecules nematic liquid crystals such as pentylcyanobiphenyl (5CB) align parallel to a bare smooth silica glass surface [43], but when the surface is functionalized with alkylsilane self-assembled monolayers formed by octadecyl- or hexyltrichlorosilane, the alignment often changes to homeotropic, depending on the degree of surface coverage [44].

Consider two types of colloidal spheres placed in a LC that are made of the same material but differ in the type of surface anchoring, one producing a perpendicular alignment, Figure 1(a), and the other a tangential alignment, Figure 1(b). The difference in surface anchoring results in two types of director distortions around the spheres [46]. Perpendicular anchoring leads to director distortions of a dipolar type with the radial-like director around the sphere being supplemented with a point defect, the so-called hyperbolic hedgehog, Figure 1(a). We direct the structural dipole $\hat{\mathbf{p}}$ from the hedgehog towards the sphere. Tangential anchoring results in quadrupolar distortions, with two points defects-boojums at the poles of the sphere, Figure 1(b).

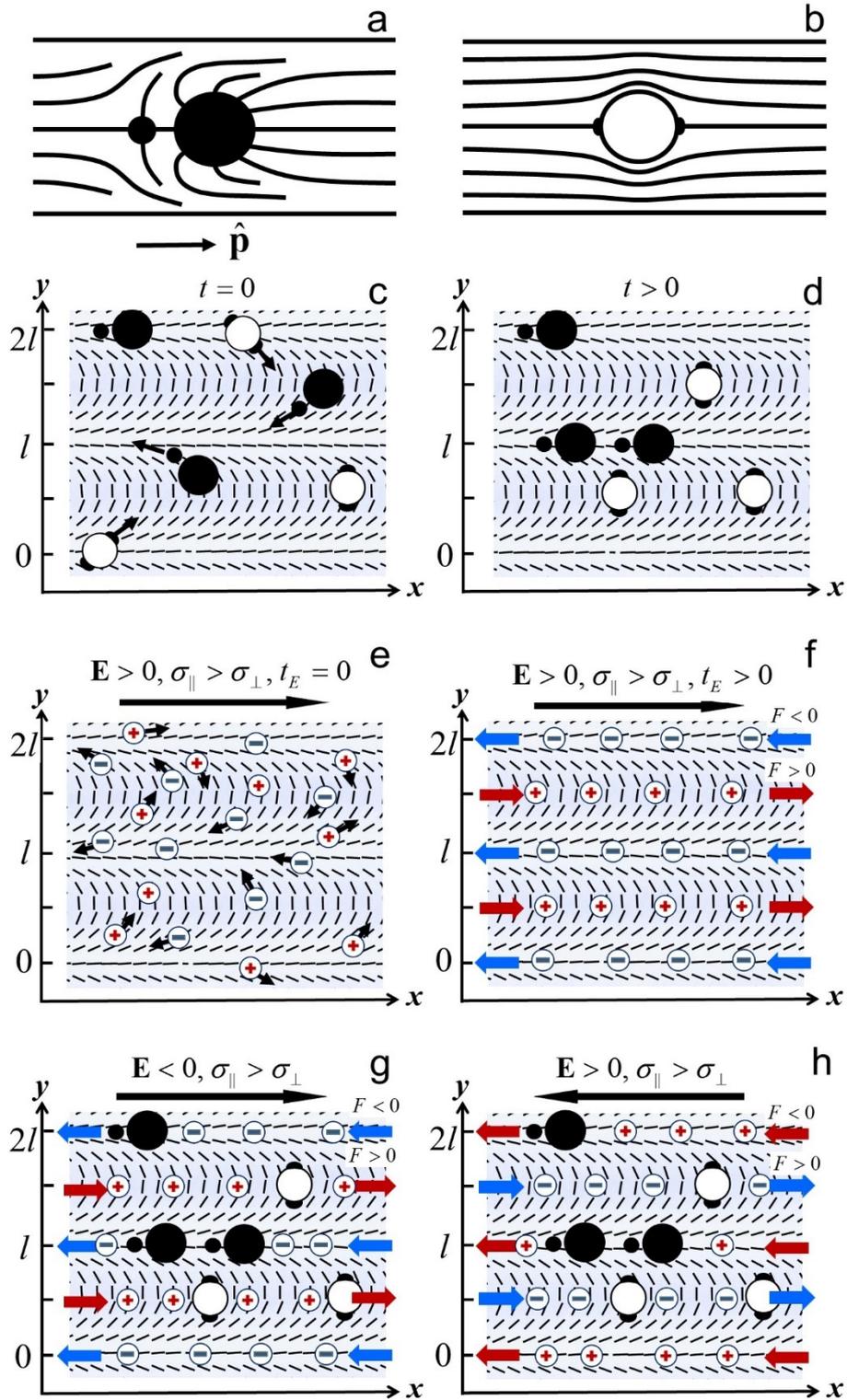

**Figure 1.** Scheme of particle sorting and separation. (a) A sphere with perpendicular anchoring creates dipolar distortions in the surrounding nematic. (b) a tangentially anchored sphere produces quadrupolar distortions. (c) Pediodic splay and bend impose elastic forces onto the particles and (d) attract the perpendicularly anchored spheres to the regions of splay and the tangentially anchored spheres to the bend regions. (e) Randomly distributed ions in the nematic bulk are moved

along the x-axis by the applied electric field but also shift along the y-axis because of anisotropic conductivity. (f) The electric field acting on the space charge produces alternating electro-osmotic flows directed along the x-axis for the bend regions (red arrows) and in the opposite direction (blue arrows) for the splay regions. (g) Combined action of elastic and electro-osmitic effects separates the particles by moving the parpendicularly anchored spheres to the left and tangentially anchored spheres to the right. (h) Reversal of the field polarity changes the sign of space charge but does not change the direction of electro-osmotic flows and the direction of particles transport.

The difference in the local director field is sufficient to separate the spheres in space at least along one spatial dimension if one uses a sandwich-like nematic cell bounded by two glass plates in which the in-plane director experiences periodic one-dimensional distortions with alternating splay and bend deformations along the $y$-axis:

$$\hat{\mathbf{n}} = (n_x, n_y, 0) = (\cos\alpha, \sin\alpha, 0), \tag{1}$$

where $\alpha = \pi y / l$, $l$ is the period (equal 80 μm in the experiments below), which is much larger than the radii $R$ of the spheres, Figure 1(c-h). The periodic pattern is created by a photoaligning technique based on plasmonic photomasks [38, 47].

Because of the long-range orientational order, the predesigned pattern in Figure 1 interacts elastically with the director fields surrounding the colloidal spheres [46]. As expected theoretically [48] and demonstrated experimentally [39], the splay regions $y=0,\pm l,\pm 2l,...$ attract colloids with the dipolar director while the bend regions $y=\pm l/2,\pm 3l/2,...$ attract the tangentially anchored spheres, Figure 1(c,d). As a result of elastic interactions between the local director distortions around the particles and the background splay-bend landscape in the photo-patterned cell, the particles with different anchoring are separated along the y-axis. Sorting along the $y$-axis is only the first stage of their complete separation. The second stage is through LCEO flows that transport particles located in the splay and bend regions towards the opposite ends of the $x$-axis. LCEO flows occur in the distorted LC acted upon by an alternating current (AC) electric field [36].

Figure 1(e,f) illustrate the underlying mechanism for the simple case when the patterned nematic cell contains no colloids and the field $\mathbf{E}_0 = (E_0, 0, 0)$ is applied along the $x$-axis.

The physical reason for LCEO is the anisotropy of the LC properties. Consider a LC with positive anisotropy of conductivity, $\Delta\sigma = \sigma_\parallel - \sigma_\perp > 0$, which means that electric carriers, ions, always present in nematics, prefer to move along the director rather than perpendicularly to it. We neglect the dielectric anisotropy for the moment. The electric field $\mathbf{E}_0 = (E_0, 0, 0)$, $E_0 > 0$ applied along the stripes, would result in electric currents not only along the $x$-axis but also along the $y$-axis, because of the conductivity anisotropy; the shifts of ions are shown by arrows in Figure 1(e). As a result, for $E_0 > 0$, negative ions accumulate at the splay stripes, $y = 0, \pm l, \pm 2l, ...$, while positive ones in the bend regions, $y = \pm l/2, \pm 3l/2, ...$, Figure 1(e,f). The field-induced spatial charge of density $\rho \propto \Delta\sigma E_0 \cos(2\pi y/l)$ experiences an electrostatic force $F = \rho E_0 \propto \Delta\sigma E_0^2 \cos(2\pi y/l)$ that drives electro-osmotic flows along the $x$-axis in the regions of bend, $y = \pm l/2, \pm 3l/2, ...$, where $F > 0$, and along the opposite direction in the splay regions, $y = 0, \pm l, \pm 2l, ...$, where $F < 0$, Figure 1(f).

The combined elastic and LCEO action of the patterned director in presence of the electric field allows one to separate the colloidal particles according to their surface properties, Figure 1(g,h). Note that the electrostatic force $F$, being proportional to the square of the field, does not change its polarity when the field polarity changes. Therefore, the electro-osmotic flows in Figure 1(g,h) preserve their directionality regardless of the polarity of the applied field; this enables one to use an AC driving for the separation. The velocities of particles driven by LCEO flows is linearly proportional to the square of the field, Figure 2.

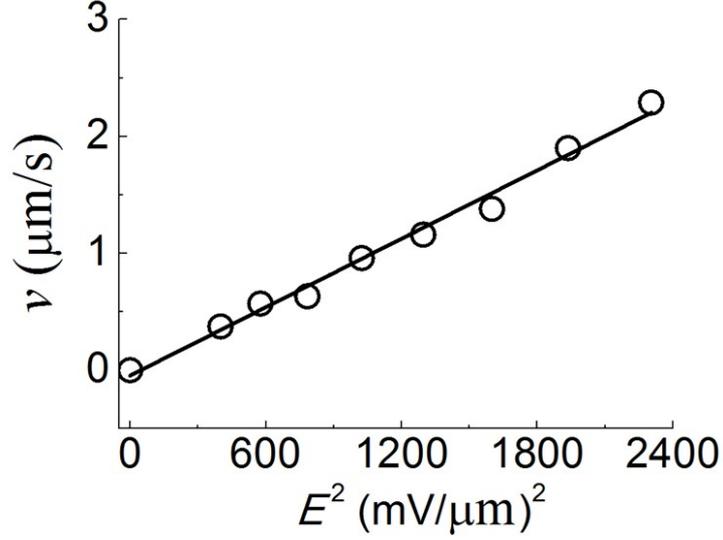

**Figure 2.** Velocity dependence of homeotropically anchored particle on the electric field. The velocity of transported particles driven by the LCEO flows grows as $E^2$.

The described splay-bend patterned director field exhibit dual functionality: (i) it elastically attracts the colloids towards the splay and bend regions and (ii) it causes charge separation and LCEO flows powered by the electric field. This dual functionality allows one to sort the colloidal particles along the $y$-axis in accordance with their surface properties and then to move them into opposite directions along the $x$-axis by LCEO, Figure 1(g,h). Below we present various examples of experimental verification of the proposed separation approach. In all experiments, we use a nematic LC mixture with zero dielectric anisotropy ($|\Delta\varepsilon| \leq 10^{-3}$) and positive anisotropy of electric conductivity, $\Delta\sigma/\bar{\sigma} = 2(\sigma_\parallel - \sigma_\perp)/(\sigma_\parallel + \sigma_\perp) \approx 0.3$, comprised of MLC7026-000 and E7 in weight proportion 89.1:10.9 (both LCs are from *EM Industries*). The cell thickness was $h = 20$ μm; all experiments are performed at room temperature 22°C. The volume fraction of colloids was about 0.1%-0.01%. The AC electric field ($E = 40$ mV/μm, frequency 5 Hz) is applied in the plane of the sample along the $x$-axis, using two indium tin oxide stripe electrodes separated by a distance 10 mm[36]. The director patterns shown in Figure 3,4 were obtained by using LC PolScope observations [49, 50].

## 3.2. Sorting and separation of polystyrene spheres with different surface functionalization and shapes

The polystyrene spheres of diameter $2R = 5$ μm are purchased from *Duke Scientific*. When their surface is treated with octadecyl-dimethyl-(3-trimethoxysilylpropyl) ammonium chloride (DMOAP), the anchoring becomes perpendicular (or homeotropic) and the director distortions acquire dipolar symmetry [37], Figure 1(a). When the surface of the same spheres is coated with streptavidin which is protein purified from bacterium *Streptomyces avidinii*, the surface alignment of the director is tangential, Figure 1(b).

When placed in the nematic cell with the director pattern, Figure 1, the streptavidin-coated polystyrene spheres are attracted to the regions of bend, while the DMOAP-coated sphere follow the splay regions, Figure 3(a). Under an applied AC electric field, the two types of spheres are separated by transport to the opposite ends of the cell, with the DMOAP-coated spheres moving to the left and the streptavidin-coated spheres moving to the right, Figure 3(a-c), See also Movie S1. Note that the DMOAP spheres move parallel to the local director, experiencing an effective viscosity $\eta_{\parallel}$, while the streptavidin spheres with tangential anchoring move perpendicularly to the local director, with a higher effective viscosity, $\eta_{\perp} > \eta_{\parallel}$. This difference contributes to somewhat slower velocities of the quadrupolar particles as compared to their dipolar counterparts. Since streptavidin has an extraordinarily high affinity for biotin, the approach demonstrated here can be potentially used to sort biotin or antibodies [51]. Furthermore, the approach can be used to separate particles that differ in shape. Figure 3(d-f) shows a boomerang-shaped particles that are located in the bend regions [52], being separated from the DMOAP-functionalized polystyrene spheres, See also Movie S2.

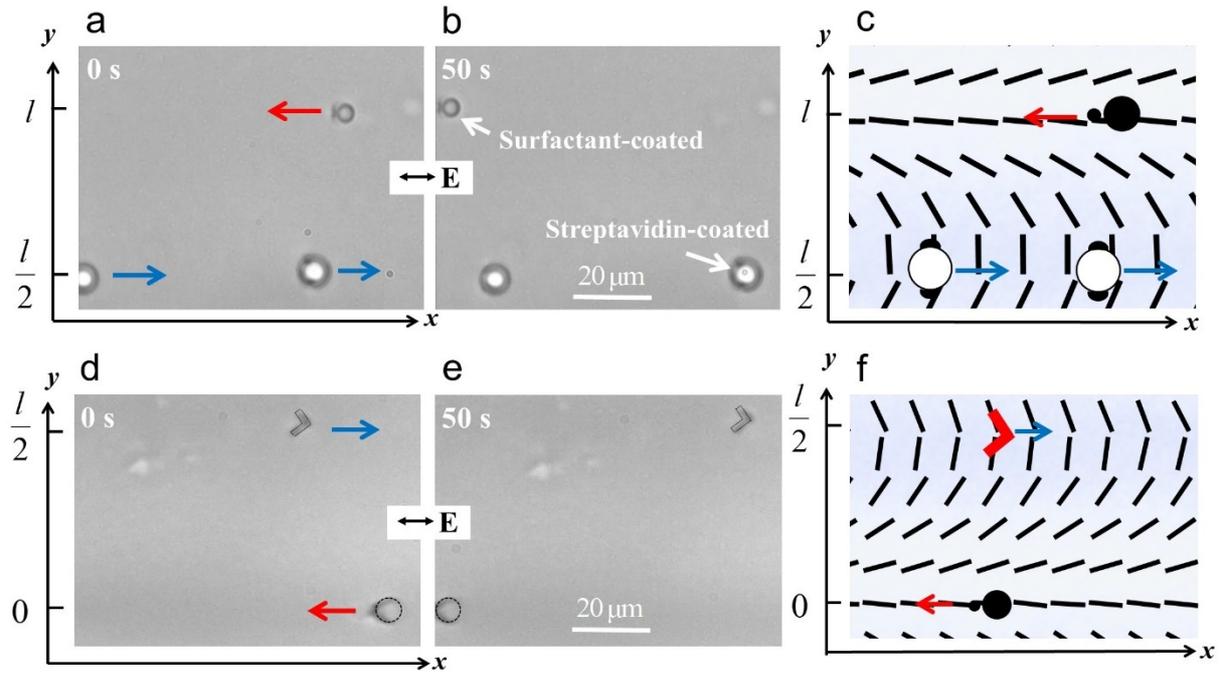

**Figure 3.** Sorting of solid particles with different surface properties by LCEO flows in patterned nematic cells. (a-c) Optical microscope images of polystyrene spheres coated with surfactant and with streptavidin sorted by the applied AC electric field ($E = 40$ mV/μm, frequency 5 Hz); time difference between the frames (a) and (b) is 50 s. The homeotropically anchored colloids in the splay region move to the left, while streptavidin-coated spheres in the bend regions move to the right (c). See also Movie S1. (d-f) Separation of spherical polystyrene particles and boomerang shaped particles. See also Movie S2. Panels c and f show the director field determined through LC PolScope observations.

### 3.3. Sorting and separation of fluid droplets with surfactant

The patterned liquid crystal-based approach to sorting and separation is also applicable to fluid droplets. We used glycerol droplets that impose a tangential anchoring of the adjacent LC [42] and glycerol with added surfactant sodium dodecyl sulfate (SDS) that sets perpendicular anchoring. The droplets of pure glycerine are attracted to the bend regions and move to the right as the electric field is applied; the homeotropically anchored glycerol droplets with surfactant move to the left, Figure 4(a-c), See also Movie S3.

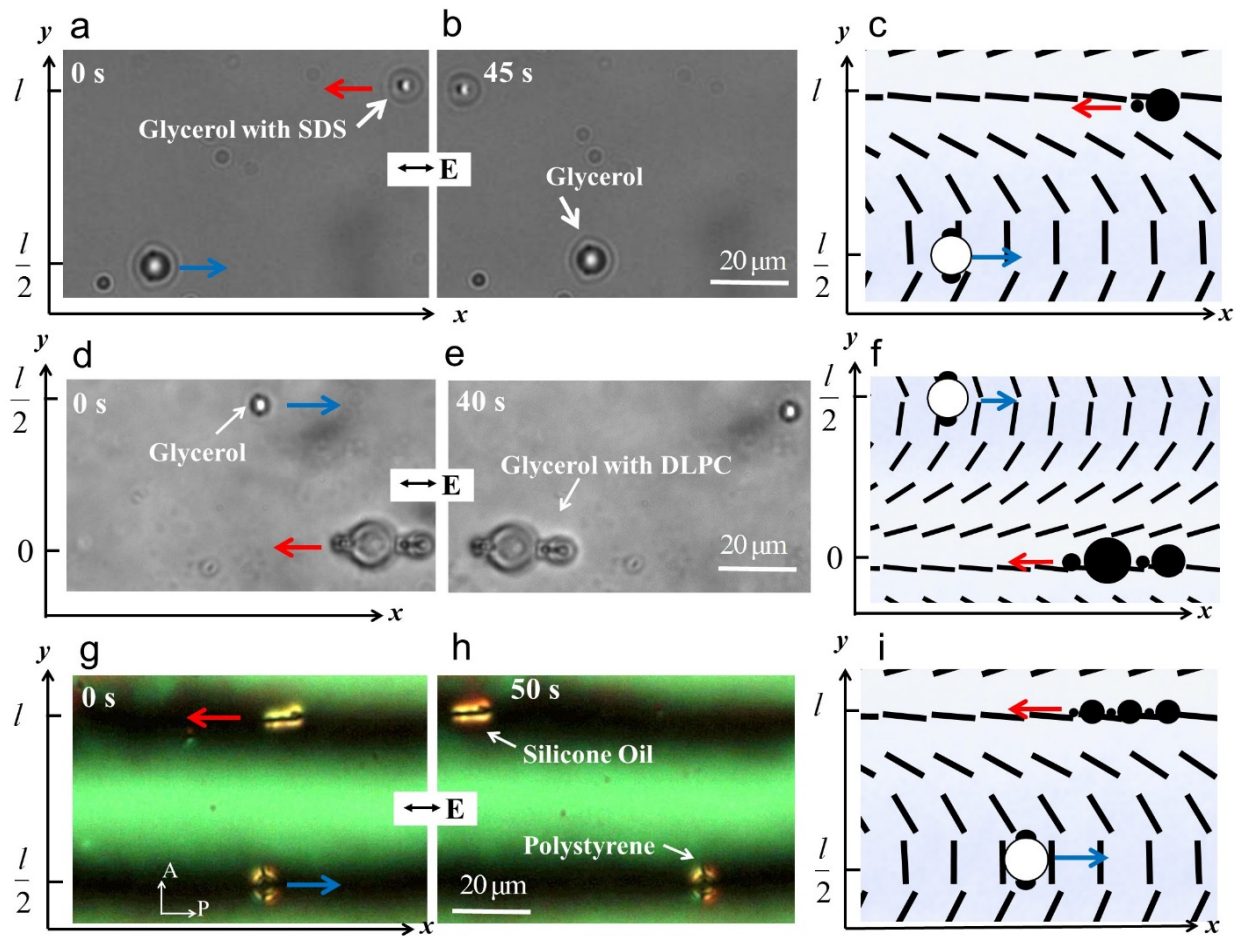

**Figure 4.** Sorting of fluid droplets with different surface properties by LCEO flows in patterned nematic cells. (a-c) Optical microscope images of separation of droplets of pure glycerol and glycerol with SDS by the applied AC electric field ( $E = 40 \text{ mV}/\mu\text{m}$ , frequency 5 Hz); the time difference between the frames (a) and (c) is 45 s. The glycerol droplets with SDS in the splay region move to the left, while the pure glycerol droplets in the bend regions move to the right. See also Movie S3. (d-f) Separation of droplets of pure glycerol and glycerol with DLPC by the applied AC electric field; the time difference between the frames (d) and (e) is 40 s. See also Movie S4. (g-i) Separation of silicone oil droplets and polystyrene spheres; microscopy observation between crossed polarizer (P) and analyzer (A). See also Movie S5. Panels c, f and I show the director field determined through LC PolScope observations.

*3.4. Sorting and separation of fluid droplets with biomolecules*

Anchoring transition on the interface between liquid crystal droplets and water has been used by Abbott et al. as a powerful approach to chemical and biological sensing [53]. Here we use glycerol droplets with added phospholipids of the type found in biological membranes, namely 1,2-Dilauroyl-sn-glycero-3-phosphorylcholine (DLPC). The glycerol droplets containing DLPC are dispersed in the liquid crystal medium. DLPC molecules settle at the glycerol-LC interface and establish a homeotropic surface anchoring of the director at it [54], Figure 4(d). As a result, DLPC-containing glycerol droplets are attracted to the splay regions and move to the left when an AC electric field is applied, while the droplets of pure glycerine are attracted to the bend regions and move to the right, Figure 4(d-f), see also Movie S4.

*3.5. Sorting and separation of fluid droplets and solid particles*

Finally, the approach allows one to separate particles of different bulk properties, provided these particles impose different orientation of the surrounding LC. For example, Figure 4(g-i) shows sorting and separation of silicone oil (purchased from *Sigma*) droplets that set homeotropic anchoring [55, 56] and polystyrene spheres with tangential anchoring, See also Movie S5.

**Conclusions**

To summarize, we propose a new technique to sort solid particles or fluid droplets with different surface properties and different shapes. The technique employs a patterned nematic LC with periodic splay and bend deformations as a dispersive medium. These patterns with the period larger than the colloidal size guide the spatial placement of solid particles or fluid droplets with different surface anchoring and different shapes. Namely, spherical particles with tangential anchoring are attracted to the regions of bend, while spheres and fluid droplets with perpendicular alignment are attracted to the regions of splay. In presence of an AC electric field, the splay-bend patterns enable electro-osmotic flows. The splay and bend stripes carry spatial electric charge of opposite signs thus the flows in these two types of stripes are antiparallel to each other. The electro-osmotic flows transport particles of different surface properties to the different ends of the sorting device, thus completing the sorting and separation task. One can wonder whether an application of the electric field could deform the director around the colloidal inclusions so that their topological properties are modified. In principle, field induced deformations are possible, caused by dielectric, conductivity, flexoelectric or surface polarization effects. However, the associated electric fields capable of noticeable changes are usually on the order of 10 V/µm, see, for example, [57, 58], which is much higher than the typical electric field needed to cause electro-osmotic flows, (10-100) mV/µm, Figure 2. Thus, the electrically powered method of particle sorting and separation should be robust against the effects caused by director coupling to the applied field. The proposed particle-sorting approach opens the opportunities in the lab-on-a-chip, cell sorting, and bio-sensing applications.

**Acknowledgements**

This work was supported by NSF grants DMR-1507637, DMS-1729509 and CMMI-1436565.

**References**

1. Sajeesh P, Sen AK. Particle separation and sorting in microfluidic devices: a review. Microfluid Nanofluid. 2014;17(1):1-52.
2. Lenshof A, Laurell T. Continuous separation of cells and particles in microfluidic systems. Chem Soc Rev. 2010;39(3):1203-1217.
3. Ramos A, García-Sánchez P, Morgan H. AC electrokinetics of conducting microparticles: A review. Curr Opin Colloid In. 2016;24:79-90.
4. Matthias S, Muller F. Asymmetric pores in a silicon membrane acting as massively parallel brownian ratchets. Nature. 2003;424(6944):53-57.
5. Huang LR, Cox EC, Austin RH, et al. Continuous particle separation through deterministic lateral displacement. Science. 2004;304(5673):987-990.
6. Yamada M, Seki M. Hydrodynamic filtration for on-chip particle concentration and classification utilizing microfluidics. Lab Chip. 2005;5(11):1233-1239.
7. Di Carlo D, Irimia D, Tompkins RG, et al. Continuous inertial focusing, ordering, and separation of particles in microchannels. P Natl Acad Sci USA. 2007;104(48):18892-18897.
8. Tierno P, Soba A, Johansen TH, et al. Dynamic colloidal sorting on a magnetic bubble lattice. Appl Phys Lett. 2008;93(21):214102.
9. Gijs MAM. Magnetic bead handling on-chip: new opportunities for analytical applications. Microfluid Nanofluid. 2004;1(1):22-40.
10. Lou XH, Qian JR, Xiao Y, et al. Micromagnetic selection of aptamers in microfluidic channels. P Natl Acad Sci USA. 2009;106(9):2989-2994.
11. Zhou H, White LR, Tilton RD. Lateral separation of colloids or cells by dielectrophoresis augmented by AC electroosmosis. J Colloid Interf Sci. 2005;285(1):179-191.
12. Velasco V, Williams SJ. Electrokinetic concentration, patterning, and sorting of colloids with thin film heaters. J Colloid Interf Sci. 2013;394:598-603.
13. Cheng IF, Chang HC, Hou D, et al. An integrated dielectrophoretic chip for continuous bioparticle filtering, focusing, sorting, trapping, and detecting. Biomicrofluidics. 2007;1(2):021503.
14. MacDonald MP, Spalding GC, Dholakia K. Microfluidic sorting in an optical lattice [10.1038/nature02144]. Nature. 2003;426(6965):421-424.
15. Applegate RW, Squier J, Vestad T, et al. Optical trapping, manipulation, and sorting of cells and colloids in microfluidic systems with diode laser bars. Opt Express. 2004;12(19):4390-4398.

16. Jonas A, Zemanek P. Light at work: The use of optical forces for particle manipulation, sorting, and analysis. Electrophoresis. 2008;29(24):4813-4851.
17. Xiao K, Grier DG. Sorting colloidal particles into multiple channels with optical forces: Prismatic optical fractionation. Phys Rev E. 2010;82(5):051407.
18. Wang ZC, Zhe JA. Recent advances in particle and droplet manipulation for lab-on-a-chip devices based on surface acoustic waves. Lab Chip. 2011;11(7):1280-1285.
19. Drinkwater BW. Dynamic-field devices for the ultrasonic manipulation of microparticles. Lab Chip. 2016;16(13):2360-2375.
20. Tasoglu S, Khoory JA, Tekin HC, et al. Levitational Image Cytometry with Temporal Resolution. Adv Mater. 2015;27(26):3901-3908.
21. Taton TA, Mirkin CA, Letsinger RL. Scanometric DNA array detection with nanoparticle probes. Science. 2000;289(5485):1757-1760.
22. Conde J, Dias JT, Grazu V, et al. Revisiting 30 years of biofunctionalization and surface chemistry of inorganic nanoparticles for nanomedicine. Front Chem. 2014;2:48.
23. Sapsford KE, Algar WR, Berti L, et al. Functionalizing Nanoparticles with Biological Molecules: Developing Chemistries that Facilitate Nanotechnology. Chem Rev. 2013;113(3):1904-2074.
24. Busseron E, Ruff Y, Moulin E, et al. Supramolecular self-assemblies as functional nanomaterials. Nanoscale. 2013;5(16):7098-7140.
25. Miller DS, Wang XG, Abbott NL. Design of Functional Materials Based on Liquid Crystalline Droplets. Chem Mater. 2014;26(1):496-506.
26. Poulin P, Stark H, Lubensky TC, et al. Novel Colloidal Interactions in Anisotropic Fluids. Science. 1997;275:1770-1773.
27. Voloschenko D, Pishnyak OP, Shiyanovskii SV, et al. Effect of director distortions on morphologies of phase separation in liquid crystals. Phys Rev E Stat Nonlin Soft Matter Phys. 2002;65(6):060701.
28. Senyuk B, Liu Q, He S, et al. Topological colloids. Nature. 2012:1-6.
29. Luo Y, Serra F, Beller DA, et al. Around the corner: Colloidal assembly and wiring in groovy nematic cells. Phys Rev E. 2016;93(3):032705.
30. Luo Y, Serra F, Stebe KJ. Experimental realization of the "lock-and-key" mechanism in liquid crystals. Soft Matter. 2016;12(28):6027-6032.
31. Muševič I. Liquid Crystal Colloids. Cham, Switzerland: Springer; 2017.
32. Wang XG, Miller DS, Bukusoglu E, et al. Topological defects in liquid crystals as templates for molecular self-assembly. Nat Mater. 2016;15(1):106-112.
33. Wang X, Kim YK, Bukusoglu E, et al. Experimental Insights into the Nanostructure of the Cores of Topological Defects in Liquid Crystals. Phys Rev Lett. 2016;116(14):147801.
34. Chen K, Gebhardt OJ, Devendra R, et al. Colloidal transport within nematic liquid crystals with arrays of obstacles. Soft Matter. 2018;14(1):83-91.
35. Lavrentovich OD, Lazo I, Pishnyak OP. Nonlinear electrophoresis of dielectric and metal spheres in a nematic liquid crystal. Nature. 2010;467:947-950.
36. Peng C, Guo Y, Conklin C, et al. Liquid crystals with patterned molecular orientation as an electrolytic active medium. Phys Rev E Stat Nonlin Soft Matter Phys. 2015;92(5):052502.
37. Lazo I, Peng C, Xiang J, et al. Liquid crystal-enabled electro-osmosis through spatial charge separation in distorted regions as a novel mechanism of electrokinetics. Nat Commun. 2014;5:5033.
38. Guo Y, Jiang M, Peng C, et al. High-Resolution and High-Throughput Plasmonic Photopatterning of Complex Molecular Orientations in Liquid Crystals Adv Mater. 2016;28:2353-2358.
39. Peng C, Turiv T, Guo Y, et al. Control of colloidal placement by modulated molecular orientation in nematic cells. Science Advances 2016;2:e1600932.
40. Chakrabarty A, Konya A, Wang F, et al. Brownian Motion of Boomerang Colloidal Particles. Phys Rev Lett. 2013;111(16):160603.
41. Sonin AA. The Surface Physics of Liquid Crystals. Gordon and Breach Publishers; 1995.
42. Volovik GE, Lavrentovich OD. The Topological Dynamics of Defects - Boojums in Nematic Drops. Zh Eksp Teor Fiz+. 1983;85(6):1997-2010.
43. Roscioni OM, Muccioli L, Della Valle RG, et al. Predicting the Anchoring of Liquid Crystals at a Solid Surface: 5-Cyanobiphenyl on Cristobalite and Glassy Silica Surfaces of Increasing Roughness. Langmuir. 2013;29(28):8950-8958.
44. Roscioni OM, Muccioli L, Zannoni C. Predicting the Conditions for Homeotropic Anchoring of Liquid Crystals at a Soft Surface. 4-n-Pentyl-4'-cyanobiphenyl on Alkylsilane Self-Assembled Monolayers. Acs Appl Mater Inter. 2017;9(13):11993-12002.
45. Pizzirusso A, Berardi R, Muccioli L, et al. Predicting surface anchoring: molecular organization across a thin film of 5CB liquid crystal on silicon. Chem Sci. 2012;3(2):573-579.

46. Poulin P, Stark H, Lubensky TC, et al. Novel colloidal interactions in anisotropic fluids. Science. 1997;275(5307):1770-1773.
47. Peng C, Turiv T, Guo Y, et al. Command of active matter by topological defects and patterns [10.1126/science.aah6936]. Science. 2016;354(6314):882.
48. Pergamenshchik VM. Elastic multipoles in the field of the nematic director distortions. Eur Phys J E. 2014;37(12):1-15.
49. Shribak M, Oldenbourg R. Techniques for Fast and Sensitive Measurements of Two-Dimensional Birefringence Distributions. Applied Optics. 2003;42(16):3009-3017.
50. Lavrentovich OD, in *Multi-scale and high-contrast PDE: from modelling, to mathematical analysis, to inversion*, H. Ammari, Y. Capdeboscq, H. Kang, Eds. (American Mathematical Society, Providence (RI), 2012), vol. 577, pp. 25-46.
51. Sarkar A, Hou HW, Mahan AE, et al. Multiplexed Affinity-Based Separation of Proteins and Cells Using Inertial Microfluidics [Article]. Scientific Reports. 2016;6:23589.
52. Peng C, Turiv T, Zhang R, et al. Controlling placement of nonspherical (boomerang) colloids in nematic cells with photopatterned director. Journal of Physics: Condensed Matter. 2017;29(1):014005.
53. Carlton RJ, Hunter JT, Miller DS, et al. Chemical and biological sensing using liquid crystals. Liq Cryst Rev. 2013;1(1):29-51.
54. Lin I-H, Miller DS, Bertics PJ, et al. Endotoxin-Induced Structural Transformations in Liquid Crystalline Droplets. Science. 2011;332(6035):1297-1300.
55. Loudet JC, Barois P, Poulin P. Colloidal ordering from phase separation in a liquid-crystalline continuous phase. Nature. 2000;407(6804):611-613.
56. Loudet JC, Richard H, Sigaud G, et al. Nonaqueous Liquid Crystal Emulsions [doi: 10.1021/la0000116]. Langmuir. 2000;16(16):6724-6730.
57. Skacej G, Zannoni C. Controlling surface defect valence in colloids. Phys Rev Lett. 2008;100(19):197802.
58. Chiccoli C, Pasini P, Skacej G, et al. Dynamical and field effects in polymer-dispersed liquid crystals: Monte Carlo simulations of NMR spectra. Phys Rev E. 2000;62(3):3766-3774.

# Supplementary materials

**Sorting and separation of microparticles by surface properties using liquid crystal-enabled electro-osmosis**

Chenhui Peng[a,1], Taras Turiv[a], Yubing Guo[a,2],

Qi-Huo Wei[a,b] and Oleg D. Lavrentovich[a,b,*]

[a]*Liquid Crystal Institute and Chemical Physics Interdisciplinary Program, Kent State Universtity, Kent, OH 44242*

[1]*Current affiliation: Research Laboratory of Electronics, Massachusetts Institute of Technology, Cambridge, MA, 02139*

[2]*Current affiliation: Max Planck Institute for Intelligent Systems, Stuttgart, Germany 70569*

*Corresponding author: e-mail: olavrent@kent.edu, tel.: +1-330-672-4844.

**Supplementary movies available online:** https://www.tandfonline.com

**Movie S1**: Polystyrene spheres coated with surfactant and with streptavidin sorted by the applied AC electric field. The homeotropically anchored colloids in the splay region move to the left, while streptavidin-coated spheres in the bend regions move to the right. The video is recorded at 1 fps; playback rate is 5 fps.

**Movie S2**: Separation of spherical polystyrene particles and boomerang shaped particles. The spherical colloids in the splay region move to the left, while boomerang shaped particles in the bend regions move to the right. The video is recorded at 1 fps; playback rate is 5 fps.

**Movie S3**: Separation of droplets of pure glycerol and glycerol with SDS by the applied AC electric field. The glycerol droplets with SDS in the splay region move to the left, while the pure glycerol droplets in the bend regions move to the right. The video is recorded at 1 fps; playback rate 10 fps.

**Movie S4**: Separation of droplets of pure glycerol and glycerol with DLPC by the applied AC electric field. The DLPC glycerol droplets in the splay region move to the left, while the pure glycerol droplets in the bend regions move to the right. The video is recorded at 1 fps; playback rate 5 fps.

**Movie S5**: Separation of silicone oil droplets and polystyrene spheres; Microscopy observation in polarized light. Silicone oil droplets set homeotropic anchoring on the interface with liquid crystal so that the silicone oil droplets in the splay region move to the left, while the polystyrene solid colloids in the bend regions move to the right. The video is recorded at 1 fps and played back at 5 fps.